\documentclass[aps,preprint,preprintnumbers,amsmath,amssymb,nofootinbib]{revtex4}
\usepackage{amssymb}
\usepackage{lscape}
\usepackage{amsmath}
\usepackage{bm}
\usepackage{graphicx}
\usepackage{epsfig}
\usepackage{color}
\begin{document}

\title{Collapse driven by a scalar field without final singularity}
\author{$^{1}$ Jaime M. Hern\'andez\footnote{E-mail: quijotecuantico@hotmail.com}, $^{2,3}$ Mauricio Bellini
\footnote{{\bf Corresponding author}: mbellini@mdp.edu.ar}, $^{1}$ Claudia Moreno\footnote{E-mail: claudia.moreno@cucei.udg.mx} }
\address{$^1$ Departamento de F\'{\i}sica, Centro Universitario de Ciencias Exactas e Ingenier\'{\i}as, Universidad de Guadalajara,
Av. Revoluci\'on 1500, Colonia Ol\'impica C.P. 44430, Guadalajara, Jalisco, M\'exico. \\
$^2$ Departamento de F\'{\i}sica, Facultad de Ciencias Exactas y
Naturales, Universidad Nacional de Mar del Plata, Funes 3350, C.P. 7600, Mar del Plata, Argentina.\\
$^3$ Instituto de Investigaciones F\'{\i}sicas de Mar del Plata (IFIMAR), \\
Consejo Nacional de Investigaciones Cient\'ificas y T\'ecnicas
(CONICET), Mar del Plata, Argentina.}

\begin{abstract}
We explore a collapsing cosmology driven by a scalar field which is minimally coupled to gravity in a spatially flat and spherically symmetric, isotropic and homogeneous space-time, with a variable timescale that avoid the final singularity. The equation of state that describes the
collapse is $\omega=1$. We calculate the back-reaction of the space-time during the collapse and the energy density fluctuations related to this back-reaction has a spectral index $n_s=0$, favouring short-wavelengths modes to be detected. The interesting is that the amplitude of these fluctuations increase with time when the collapse is sufficiently strong.
\end{abstract}
\maketitle

\section{Introduction and motivation}

Moreover, the seminal work of Oppenheimer and Snyder \cite{os}, the study of the evolution of a scalar field is of particular interest in a gravitational collapse as well as cosmological contexts when the spherical symmetry is preserved. One of the major mechanisms for which scalar fields are thought to be responsible is the inflationary scenario \cite{infl}, or quintessential cosmological models \cite{st}, where the pressure $P$, related to some scalar field $\phi$ is negative, and governed by the potential energy density: $P={\dot\phi^2\over 2} -V(\phi)$. Furthermore, a scalar field with a variety of potential can mimic the evolution of many kinds of matter with spherically symmetric distributions \cite{moss}. Many exact solutions were found to describe the collapse \cite{ch}, but it is not known how the final singularity can be avoided in the final state when we describe the dynamical evolution of a gravitational collapse.

On the other hand, it is expected that during the collapse of a star, or some spherically symmetric region of the universe, the space-time back-reaction suffered by the space-time is important enough to be detected, at least at sufficiently small distances, because, due to the sphericity of the source (i.e., the collapsing system), the wave emitted must be a spherical wave of space-time. These back-reaction effects cannot be considered in a linear approximation when we are in presence of strong gravitational fields \cite{rb}. Therefore we must use a non-perturbative formalism in which these effects can be described without approximations.

The aim of this work is to describe dynamically a collapse driven by a scalar field that can avoid the final singularity. To do it we must consider a co-moving system using a Friedmann-Robertson-Walker (FRW) metric, but with a variable time-scale, which will be responsible for describing the relativistic temporal evolution of a co-moving clock with the collapse. The relativistic behavior of the time should very important to describe the breaking of the collapse when it is in its final stage. 

The paper is organized as follows: in Sect. II we develop the model of the collapse driven by a scalar field on a background metric with the variable timescale, we revisit the minimum action principle with boundary conditions incorporated and the description of these effects on the background metric. In Sect. III we describe the
model of the collapse in a particular example where the final singularity is avoided. Finally, in Sect. IV we develop some final comments.

\section{The model of collapse, with variable timescale.}

We shall suppose that the timescale of the collapse is not constant, so that, a spatially flat, isotropic and homogeneous collapsing universe, can be represented by a line element
\begin{equation}\label{back}
d{S}^2 = e^{-2\int \Gamma(t)\, \,dt}dt^2 - a_0^2 \,\, e^{2\int h(t) dt}\,\, {\delta}_{ij}\, \,d{x}^i d{x}^j,
\end{equation}
which describes the background of the collapse. Here, $h(t)<0$ is the collapse rate parameter on the background metric and $\Gamma(t)$ describes the time scale of the background metric. In this paper, we shall consider natural units so that $c=\hbar=1$. In order to describe a collapse, we shall consider the action for a scalar field $\phi$ which is minimally coupled to gravity\footnote{We shall denote with a {\em hat}, geometrical quantities calculated on the Riemann (background) manifold.}.
If its dynamics is governed by a scalar potential $V(\phi)$, the action can be written as
\begin{equation}\label{1}
{\cal I} = \int d^4x \, \sqrt{-\hat{g}} \,\left\{ \frac{{ \hat{R}}}{16\pi G} - \left[\frac{\dot\phi^2}{2}\,e^{2\int \Gamma(t)\,dt} - V(\phi)\right]\right\},
\end{equation}
where the volume of the background manifold is $\hat{v}=\sqrt{-\hat{g}}=a^3_0\,e^{-\int \Gamma(t) \,dt}\,e^{3\int H(t)\, \,dt}$.
The action (\ref{1}) can be rewritten as
\begin{equation}\label{2}
{\cal I} = \int d^4x \, \sqrt{-\hat{g}}\,e^{2\int \Gamma(t)\,dt}\,\left\{ \frac{ \bar{R}}{16\pi G} - \left[\frac{\dot\phi^2}{2} - \bar{V}(\phi)\right]\right\},
\end{equation}
that can be considered as an action for a minimally coupled to gravity scalar field on a effective background volume $\hat{\bar{v}}=\sqrt{-\hat{g}}\,e^{2\int \Gamma(t)\,dt}$, a redefined potential $\bar{V}(\phi)=V(\phi)\,e^{-2\int \Gamma(t)\,dt}$, and an effective scalar curvature $\bar{R}= \hat{R}\,e^{-2\int \Gamma(t)\,dt}$.

\subsection{background dynamics}

The effective volume of the background manifold in (\ref{2}), is $\hat{\bar{v}}=\sqrt{-\hat{g}} \,e^{2\int \Gamma(t)\,dt}=a^3_0\,e^{\int \Gamma(t)\, \,dt}\,e^{3\int h(t)\, \,dt}$. The dynamics of the scalar field $\phi$ is given by
\begin{equation}
\ddot\phi + \left[3 h(t)+\Gamma(t)\right] \dot\phi + \frac{\delta \bar{V}}{\delta\phi} =0. \label{infl}
\end{equation}
The background Einstein equations, are
\begin{eqnarray}
3 h^2  &=& 8\pi\, G \,\rho, \label{a} \\
-\left[3 h^2 + 2 \dot{h} + 2 \Gamma\, h\right]  &=& 8 \pi \,G\,P, \label{b}
\end{eqnarray}
where $P= \left(\frac{\dot{\phi}^2}{2}-\bar{V}(\phi)\right)\, e^{2\int \Gamma(t)\, \,dt} $ is the pressure and $\rho=\left(\frac{\dot{\phi}^2}{2} + \bar{V}(\phi)\right)\, e^{2\int \Gamma(t)\, \,dt}$ the energy density due to the scalar field. In a collapsing system, the pressure will be negative, but the kinetic component of the energy density will be significative during the evolution of the system. The equation of state that describes the dynamics of the system is:
\begin{equation}
\omega= \frac{P}{\rho} = -\left(1+\frac{2 \dot{h}}{3 h^2}+\frac{2 \Gamma}{3 h}\right). \label{om}
\end{equation}
From the physical point of view, if we consider a co-moving frame where $U^{0}=\pm \sqrt{g^{00}}$ and $U^j=0$, such that $j$ can take the values
$j=1,2,3$, the relativistic velocity, $U^0={dx^0\over dS}$, will describe the rate of time suffered by a relativistic observer which is falling with the collapse of the system. Notice that the velocity will always hold the expression: $g_{\mu\nu}\, U^{\mu}\,U^{\nu}=1$.

\subsection{The back-reaction with variable time-scale}

If we deal with an orthogonal base,
the curvature tensor will be written in terms of the connections:
\begin{equation}R^{\alpha}_{\,\,\,\beta\gamma\delta} = \Gamma^{\alpha}_{\,\,\,\beta\delta,\gamma} -  \Gamma^{\alpha}_{\,\,\,\beta\gamma,\delta}
+ \Gamma^{\epsilon}_{\,\,\,\beta\delta} \Gamma^{\alpha}_{\,\,\,\epsilon\gamma} - \Gamma^{\epsilon}_{\,\,\,\beta\gamma}
\Gamma^{\alpha}_{\,\,\,\epsilon\delta}. 
\end{equation}
To describe exact back-reaction effects, we shall consider Relativistic Quantum Geometry (RQG), introduced in \cite{rb}. In this formalism, the manifold is defined with the connections\footnote{To simplify the notation we denote $\sigma_{\alpha} \equiv \sigma_{,\alpha}$.}
\begin{equation}\label{gama}
\Gamma^{\alpha}_{\beta\gamma} = \left\{ \begin{array}{cc}  \alpha \, \\ \beta \, \gamma  \end{array} \right\}+ \sigma^{\alpha} \hat{g}_{\beta\gamma} ,
\end{equation}
where $\delta{\Gamma^{\alpha}_{\beta\gamma}}=\sigma^{\alpha} \hat{g}_{\beta\gamma} $ takes into account the displacement of the Weylian manifold \cite{weyl} with respect to the Riemannian background, which is described by the Levi-Civita symbols in (\ref{gama}). In our approach, $\sigma(x^{\alpha})$ is a scalar field and the covariant derivative of the metric tensor in the Riemannian background manifold is null (we denote with a semicolon the Riemannian-covariant derivative): $\Delta g_{\alpha\beta}=g_{\alpha\beta;\gamma} \,dx^{\gamma}=0$. However, the Weylian covariant derivative \cite{weyl} on the manifold generated by (\ref{gama}) is nonzero: $ g_{\alpha\beta|\gamma} = \sigma_{\gamma}\,g_{\alpha\beta}$. From the action's point of view, the scalar field $\sigma(x^{\alpha})$ is a generic geometrical transformation that leaves invariant the action \cite{rb}
\begin{equation}\label{aac}
{\cal I} = \int d^4 \hat{x}\, \sqrt{-\hat{g}}\,e^{2\int \Gamma(t)\,dt}\, \left[\frac{\hat{R}}{2\kappa} + \hat{{\cal L}}\right] = \int d^4 \hat{x}\, \left[\sqrt{-\hat{g}}\,e^{2\int \Gamma(t)\,dt} e^{-2\sigma}\right]\,
\left\{\left[\frac{\hat{R}}{2\kappa} + \hat{{\cal L}}\right]\,e^{2\sigma}\right\}.
\end{equation}
Hence, Weylian quantities will be varied over these quantities in a semi-Riemannian manifold so that the dynamics of the system preserves the action: $\delta {\cal I} =0$, and we obtain
\begin{equation}
-\frac{\delta{\bar{v}}}{\hat{\bar{v}}} = \frac{\delta \left[\frac{\hat{R}}{2\kappa} + \hat{{\cal L}}\right]}{\left[\frac{\hat{R}}{2\kappa} + \hat{{\cal L}}\right]}
= 2 \,\delta\sigma,
\end{equation}
where $\delta\sigma = \sigma_{\mu} dx^{\mu}$ is an exact differential and $\hat{\bar{v}}=\sqrt{-\hat{ g}}\,e^{2\int \Gamma(t)\,dt}$ is the volume of the Riemannian manifold, such that the volume the varied system is ${\bar{v}}=\sqrt{-\hat{ g} }\,e^{2\int \Gamma(t)\,dt}\,e^{-2\sigma}$. Of course, all the variations are in the Weylian geometrical representation and assure us gauge invariance because $\delta {\cal I} =0$.
The Ricci tensor in the extended manifold $\bar{R}_{\alpha\beta}$, is related with the Ricci tensor in the Riemann manifold $\hat{R}_{\alpha\beta}$, by the expression
\begin{equation}
\bar{R}_{\alpha\beta} =\hat{R}_{\alpha\beta} +  \sigma_{\alpha ; \beta} + \sigma_{\alpha} \sigma_{\beta} - g_{\alpha\beta}
\left[ \left(\sigma^{\mu}\right)_{;\mu} + \sigma_{\mu} \sigma^{\mu} \right],
\end{equation}
so that both representations of the scalar curvature, are related by
\begin{equation}
\bar{R} = \hat{R} - 3\left[\left( \sigma^{\mu}\right)_{;\mu} + \sigma_{\mu} \sigma^{\mu} \right],
\end{equation}
where $\hat R = 6 \, p (\alpha - 2p - 1) \frac{t^{2(\alpha-1)}}{t_0^{2 \alpha}} $.

The Einstein tensor can be written as
\begin{equation}
\bar{G}_{\alpha\beta} = \hat{G}_{\alpha\beta} + \sigma_{\alpha ; \beta} + \sigma_{\alpha} \sigma_{\beta} + \frac{1}{2} \,g_{\alpha\beta}
\left[ \left(\sigma^{\mu}\right)_{;\mu} + \sigma_{\mu} \sigma^{\mu} \right],
\end{equation}
where we have made use of the fact that the connections are symmetric.

From the point of view of the Weylian-like manifold: $\Lambda\equiv \Lambda(\sigma, \sigma_{\alpha})$ can be considered a functional \cite{rb}, given by
\begin{equation}\label{aa}
\Lambda(\sigma, \sigma_{\alpha}) = -\frac{3}{4} \left[ \sigma_{\alpha} \sigma^{\alpha} + \hat{\Box} \sigma\right].
\end{equation}
Therefore, a geometrical quantum action on the Weylian-like manifold with (\ref{aa}), by defining the Lagrangian density ${\cal L} = -\frac{2}{3\kappa} \, \Lambda(\sigma, \sigma_{\alpha})$:
\begin{equation}
{\cal W} = \int d^4 x \, \sqrt{-\hat{g}}\,e^{2\int \Gamma(t)\,dt} \,\, {\cal L},
\end{equation}
such that the dynamics of the geometrical field is given by the Euler-Lagrange equations, after imposing $\delta
{\cal W}=0$. 

The dynamics of the back-reaction is described by the equation
\begin{equation}\label{si}
\ddot\sigma + \left[3h+\Gamma\right] \dot\sigma -\frac{1}{a^2_0} e^{-2\int \left[h+\Gamma\right]\,dt}\, \nabla^2 \sigma=0.
\end{equation}
The geometrical scalar field $\sigma$ is expressed in terms of a Fourier expansion
\begin{equation}\label{four}
\sigma\left(\vec{x},t\right) = \frac{1}{(2\pi)^{3/2}} \int \, d^3k \, \left[ A_k \, e^{i \vec{k}.\vec{x}} \xi_k(t) + A^{\dagger}_k \, e^{-i \vec{k}.\vec{x}} \xi^*_k(t) \right],
\end{equation}
with creation and annihilation operators $A^{\dagger}_k$ and $A_k$, that comply with the algebra
\begin{eqnarray}\label{m5}
\left<B\left|\left[A_{k},A_{k'}^{\dagger}\right]\right|B\right>&=&\delta ^{(3)}(\vec{k}-\vec{k'}), \nonumber \\
\left<B\left|\left[A_{k},A_{k'}\right]\right|B\right>&=&\left<B\left|\left[A_{k}^{\dagger},A_{k'}^{\dagger}\right]\right|B\right>=0.
\end{eqnarray}
When time-scale is variable, the metric with back-reaction effects included results to be
\begin{equation}\label{met1}
g_{\mu\nu} = {\rm diag}\left[e^{-2\int \Gamma(t)dt}\, e^{2\sigma}, - a_0^2 \,\, e^{2\int h(t) dt}\,\, e^{-2\sigma}, - a_0^2 \,\, e^{2\int h(t) dt}\,\, e^{-2\sigma}, - a_0^2 \,\, e^{2\int h(t) dt}\,\, e^{-2\sigma}\right],
\end{equation}
where the background scale factor $a(t)$ will be given by (\ref{sfactor}). Notice that its volume is described by (\ref{met1}), is $\bar{v}= \sqrt{-\hat{g}}\,e^{2\int \Gamma(t)\,dt} \,e^{-2\sigma}= a_0^3 \,e^{3\int h(t) dt}\,\,e^{\int \Gamma(t) dt}\, e^{-2\sigma}$. The relativistic quantum algebra is given by the expressions
\begin{eqnarray}\label{con}
\hat{\bar{v}}\left[\sigma(x),\sigma^{\alpha}(y) \right] &=& i \, \hat{U}^{\alpha}\, \delta^{(4)} (x-y), \nonumber \\ \hat{\bar{v}}\left[\sigma(x),\sigma_{\alpha}(y) \right] &=& -i\, \hat{U}_{\alpha}\, \delta^{(4)} (x-y),
\end{eqnarray}
where $U^0=e^{\int \Gamma(t)dt}$, $U^i=0$, are the relativistic velocities on the Riemannian (background) manifold so that they are calculated using the geodesic equation with Levi-Civita connections with $U^{\alpha}U_{\alpha}=1$.

Furthermore, as calculated in a previous work \cite{Mio}, the variation of the energy density fluctuations is
\begin{equation}\label{de}
\left<B\left|\frac{1}{\bar{\rho}} \frac{\delta \bar{\rho}}{\delta S}\right|B\right> = - 2 \frac{\delta\sigma}{\delta S}= -2 \,U^0\, \sigma_0 =  -2 \,U^0\,\dot{\sigma},
\end{equation}
where $U^0= \left(\frac{t}{t_0}\right)^{\alpha}$ and $\dot{\sigma} \equiv \left<B\left|\dot\sigma^2\right|B\right>^{1/2}$.

\section{An example}

We consider the case where the collapse rate parameter $h(t)=\dot{a}/a$ and the function that characterizes the time scale $\Gamma(t)$, are
\begin{equation}\label{ex}
\Gamma(t) = \alpha/t, \qquad h(t) = -p/t,
\end{equation}
where the scale of the system (can be the scale factor of the universe or the ratio of a collapsing sphere), is
\begin{equation}\label{sfactor}
a(t)= a_0\, \left(\frac{t}{t_0}\right)^{-p},
\end{equation}
such that $t_0$ and $a_0$ are respectively the initial values for $t$ and $a$. For $p>0$ values, it is expected that $\lim_{t\rightarrow \infty}{a(t)}\rightarrow 0$. It is expected that $p>0$, in order to $h(t)<0$, because we are dealing with a contracting system. The background dynamics are given by the Einstein equations (\ref{a},\ref{b}), with the scalar field dynamics (\ref{infl}). From the Einstein equations, we can obtain
\begin{eqnarray}
\dot{\phi}^2 &=& -\frac{1}{4\pi G} \left(\dot{h} + \Gamma h\right), \label{aa} \\
\bar{V} & = & \frac{1}{8\pi G} \left[3 h^2 + \left(\dot{h} + \Gamma h\right)\right],
\end{eqnarray}
so that, by choosing the positive root in (\ref{aa}), we obtain
\begin{equation}
\dot{\phi}(t) =  - \sqrt{\frac{p(\alpha-1)}{4\pi\, G}} \,t^{-1}.
\end{equation}

The temporal evolution for the scalar field and the potential written in terms of the scalar field, are:
\begin{eqnarray}
\phi(t) - \phi_0 &=&  \mp \frac{1}{2 \sqrt{\pi G}} \left[p (\alpha-1)\right]^{1/2}\, \ln{(t/t_0)}, \\
\bar{V}(\phi) &=& \frac{p(\alpha-1-3p)}{8\pi G t^2_0} \, e^{\mp 4\sqrt{\frac{\pi G}{p(\alpha-1)}}\,(\phi-\phi_0)}.
\end{eqnarray}

Since $\frac{\delta\bar{V}}{\delta\phi} \equiv \frac{\dot{\bar{V}}}{{\dot{\phi}}}$, using the expressions (\ref{ex}) in Eq. (\ref{infl}), we obtain the condition
\begin{equation}
\sqrt{\frac{p}{\pi\,G(\alpha-1)}} \frac{\alpha \left(\alpha-3 p  -1\right)}{2} \,t^{-2} =0. \label{ee}
\end{equation}
Since we are interested in studying a collapse with an equation of state that describes deceleration in order to the collapse can stop, we must consider values with $p>0$ in order to $h(t)<0$. The solutions for $\alpha$ are given by the roots of (\ref{ee}):
\begin{equation}
\alpha =3p+1, \label{uu}
\end{equation}
from which it is easy to see that $\alpha \geq 1$.  The equation of state that describes the collapse [see eq. (\ref{om})], is
\begin{equation}
\omega = 1.
\end{equation}
Since the time velocity in the action (\ref{1}) is given by the expression: $U^0\equiv \frac{dx^0}{ds}= \pm \sqrt{g^{00}}$ in a co-moving system, the value with $\alpha >0$ corresponds to a $U^0$ that increases with $t$, so that the time is accelerated
\begin{equation}\label{vel}
U^0= \left(\frac{t}{t_0}\right)^{(3p+1)},
\end{equation}
for $t \geq t_0$. However, after the mapping this velocity becomes unitary in the action (\ref{2}), and the background volume  $\hat{\bar{v}}=\sqrt{-\hat{g}}$\footnote{Here, $\hat{g}$ is the determinant of the covariant metric tensor.} of the Riemann manifold of this action, is:
\begin{equation}\label{vol}
\hat{\bar{v}} = a^3_0\, \left(\frac{t}{t_0}\right),
\end{equation}
which corresponds to a 4d-volume that increases with time. The equation of motion for the modes $\xi_k(t)$ in the expansion (\ref{four}), is
\begin{equation}\label{xi}
\ddot\xi_k(t) + \frac{1}{t} \dot\xi_k(t) + \frac{k^2}{a_0^2} \left(\frac{t}{t_0}\right)^{-2(2p+1)}\xi_k(t)=0.
\end{equation}
Sometimes it is useful rewrite the equation (\ref{xi}), for the redefined modes $\xi_k(t) = \psi_k(t)\,\, e^{-\frac{1}{2} \int \,t^{-1} \,dt}$,
so that
\begin{equation}\label{psi}
\ddot\psi_k(t) + \omega^2(k,t)\,\psi_k(t)=0,
\end{equation}
where the $(k,t)$-squared frequency is
\begin{equation}
\omega^2(k,t) = \left[\frac{k^2}{a_0^2} \left(\frac{t}{t_0}\right)^{-2(2p+1)} + \frac{1}{4 t^2}\right].
\end{equation}
Notice that $\omega^2(k,t) >0$, so that the system is always stable.
\begin{equation}
\lim_{t\rightarrow \infty} \omega(k,t) \rightarrow 0.
\end{equation}
Therefore, the solutions of $\xi_k(t)$ will be periodic, but with a frequency that decreases with $t$.
Using the commutation relation (\ref{m5}) and the Fourier
expansions (\ref{four}) in
\begin{equation}
\left<B\left|\left[\sigma(t,\vec{x}), \Pi_{0}(t,\vec{x}')\right]\right|B\right> = -i\,\delta^{(3)}(\vec{x}-\vec{x}'),
\end{equation}
with canonical momentum: $\Pi_{\alpha}=\frac{\delta {\cal L}_{q}}{\delta \sigma^{\alpha}}=-{3\over 4} \hat{\bar{v}} \,\sigma_{\alpha}$,
we obtain the normalization condition for the modes $\xi_{k}(\tau)$, such that $\hat{\bar{v}}$ is the volume of the manifold in the action (\ref{2})
\begin{equation}\label{m6}
\psi_{k}(t) \dot{{\psi}}^*_{k}(t) - \psi^*_{k}(t) \dot{{\psi}}_{k}(t) = -\frac{{\rm i}}{a^3_0},
\end{equation}
where the asterisk denotes the complex conjugated and $\bar{v}(t)$ is given by (\ref{vol}). The solution of Eq. (\ref{xi}), once quantised, is\footnote{To quantise the modes in the ultraviolet limit, where $y\gg 1$, we use the asymptotic expressions
\begin{displaymath}
\left.{\cal H}^{(1,2)}_{\nu}[y]\right|_{y\gg 1} \simeq \sqrt{\frac{2}{\pi\,y}} \,e^{\pm {\rm i}[y-\nu\pi/2 - \pi/4]}.
\end{displaymath} }
\begin{equation}
\xi_k \left(t \right) ={\rm i}\,\sqrt{\frac{\pi \,t}{2\,p\,a_0^3}}\,\,\,{\cal H}^{(2)}_{0}\left[y(t)\right],
\end{equation}
where ${\cal H}^{(2)}_{\nu}\left[y(t)\right]$ is the second kind Hankel function, with argument
\begin{equation}
y(k,t)=k\, \frac{t_0^{2p+1}}{2 a_0 p} \,t^{-2p}.
\end{equation}
Notice that for sufficiently large $t$: $\lim_{t\rightarrow \infty} y(k,t) \rightarrow 0$. The time derivative of the time-dependent modes, are
\begin{equation}
\dot\xi_k(t)= {\rm i}\,\sqrt{\frac{\pi }{2\,p\,a_0^3}}\,\,\left[\frac{1}{2 \sqrt{t}} \,{\cal H}^{(2)}_{0}\left[y(k,t)\right]+\frac{k}{a_0} \,{t^{\frac{1}{2}} \left(\frac{t}{t_0}\right)^{-(2\,p+1)}\,} \,{\cal H}^{(2)}_{1}\left[y(k,t)\right]\right],
\end{equation}
so that the squared norm related to $\dot\xi_k(t)$ is
\begin{eqnarray}
\dot\xi_k(t)\,\dot\xi^*_k(t)&=& \frac{\pi}{2\,p\,a_0^3}\,\left[\frac{1}{4\,t}{\cal H}^{(2)}_{0}\left[y(k,t)\right] \,{\cal H}^{(1)}_{0}\left[y(k,t)\right] +
\right. \nonumber \\
&& \left.
\frac{{k}^{2}}{a_0^2}\,{{t}}\left(\frac{t}{t_0}\right)^{-2(2p+1)}
{\cal H}^{(2)}_{1}\left[y(k,t)\right] \,{\cal H}^{(1)}_{1}\left[y(k,t)\right]\right].
\end{eqnarray}

The power spectrum of the energy density fluctuations due to back-reaction effects, are
\begin{eqnarray}
{\cal P}_{\left<B\left|\frac{1}{\bar{\rho}} \frac{\delta \bar{\rho}}{\delta S}\right|B\right>}(k,t) & = & {\frac{k^3}{4\,\pi\,p\,a_0^3}} \,\left[\frac{1}{4\,t}{\cal H}^{(2)}_{0}\left[y(k,t)\right] \,{\cal H}^{(1)}_{0}\left[y(k,t)\right] +
\right. \nonumber \\
&& \left.
\frac{{k}^{2} t_0}{a_0^2}\, \left(\frac{t}{t_0}\right)^{-p}
{\cal H}^{(2)}_{1}\left[y(k,t)\right] \,{\cal H}^{(1)}_{1}\left[y(k,t)\right]\right]. \label{po}
\end{eqnarray}
This expression is valid on all scales, if we suppose that the collapse is homogeneous on all scales. In a more realistic model for a collapse one would use a metric which describes an inhomogeneous one.

\subsection{Large-scale power spectrum of energy density fluctuations}

In order to study the large-scale spectrum, we must calculate the asymptotic limit of the product $\left.{\cal H}^{(2)}_{1}\left[y(k,t)\right] \,{\cal H}^{(1)}_{1}\left[y(k,t)\right]\right|_{y \ll 1}$. The result for this case is

\begin{equation}
\left[{\cal H}^{(2)}_{0}\left[y(k,t)\right] \,{\cal H}^{(1)}_{0}\left[y(k,t)\right]\right|_{y \ll 1} \simeq \frac{1}{ \Gamma^2(1)} + \frac{\Gamma^2(0)}{\pi^2},\label{as1}
\end{equation}
\begin{equation}
\left[{\cal H}^{(2)}_{1}\left[y(k,t)\right] \,{\cal H}^{(1)}_{1}\left[y(k,t)\right]\right|_{y \ll 1} \simeq \frac{1}{ \Gamma^2(2)} \left(\frac{y(k,t)}{2}\right)^2+ \frac{\Gamma^2(1)}{\pi^2} \left(\frac{y(k,t)}{2}\right)^{-2},\label{as}
\end{equation}
where the last term in (\ref{as}) is the dominant for $y(k,t) \ll 1$, so that we can make the approximation
\begin{equation}
\left[{\cal H}^{(2)}_{1}\left[y(k,t)\right] \,{\cal H}^{(1)}_{1}\left[y(k,t)\right]\right|_{y \ll 1} \simeq k^{-2}\, \frac{\Gamma^2(1)}{\pi^2} \, \left(\frac{a_0\,p}{t^{2p+1}_0}\right)^2 \, t^{4p},\label{as1}
\end{equation}
and therefore, the large-scale power spectrum of the energy density fluctuations due to back-reaction effects
\begin{eqnarray}
{\cal P}_{\left<B\left|\frac{1}{\bar{\rho}} \frac{\delta \bar{\rho}}{\delta S}\right|B\right>}(k,t) & \simeq &  \frac{k^3}{4\, \pi^3 \,a_0^3}\left[\frac{\pi^2}{4 \, p}\left(\frac{1}{ \Gamma^2(1)}+ \frac{\Gamma^2(0)}{\pi^2} \right)\frac{1}{t}+\frac{\Gamma^2(1)\,p}{ t_0}\left(\frac{t}{t_0}\right)^{3p}\right]. \label{po1}
\end{eqnarray}
Notice that the first term in (\ref{po1}) decreases with time and correspond to a spectral index $n_s=-2$. However, the dominant term (the second one) increases with time. Its spectral index is related to the parameter $\nu$, with the expression: $3-2\nu=1-n_s$. In our case $\nu=1$, so that $n_s=0$, which means that, for a given $t$, the amplitude for the energy density fluctuations increases on small scales (for smallest wavelengths). Therefore, the detection of these phenomena should be more efficient on high frequencies. Notice that for $p<1$, the time evolution of (\ref{po1}) are squeezed and the amplitude of ${\cal P}_{\left<B\left|\frac{1}{\bar{\rho}} \frac{\delta \bar{\rho}}{\delta S}\right|B\right>}(k,t)$ decreases with time in slow-rate-collapses. However, for a sufficiently strong-rate-collapse with $p>1$, the amplitude of ${\cal P}_{\left<B\left|\frac{1}{\bar{\rho}} \frac{\delta \bar{\rho}}{\delta S}\right|B\right>}(k,t)$ will be increased with time. This kind of collapses would be more easily detected by experiments.

\section{Final Comments}

We have studied the dynamics of a collapse spherically symmetric driven by a scalar field that avoid the final singularity jointly with the
geometrical back-reaction of space-time produced by this collapse. During the collapse, the system reaches a tiny equation of state $\omega=1$ with a spectral index: $n_s= 0$, which favors the detection on wavelengths with size $\lambda \ll a_0$, smaller than the initial size of the collapsing ball.

We got two important results the first one is that the co-moving relativistic observer never reaches the center of the sphere, because the physical time evolution
$d\tau=U_{0}\,dx^0=\sqrt{g_{00}}\,dx^0 = \left(\frac{t}{t_0}\right)^{-(3p+1)}\,dt$ [see eq. (\ref{vel})], decelerates for a co-moving observer which falls with the collapse. And the second one is that the amplitude of the energy density fluctuations increases with time for abrupt collapses. Of course, the model would be more realistic if the collapse were considered as inhomogeneous. This case will be deal in a future work.

\section*{Acknowledgements}

\noindent M. B. acknowledges CONICET, Argentina (PIP 11220150100072CO) and UNMdP (EXA852/18) for financial support. This research was supported by the CONACyT Network Project No. 294625 `Agujeros Negros y Ondas Gravitatorias".

\end{document}